\documentclass[useAMS,usenatbib]{mn2e}
\usepackage{graphicx}

\title[Remote Sounding]{Satellite characterization of four interesting sites for astronomical instrumentation}

\author[S. Cavazzani et al.]{S. Cavazzani$^{1}$\thanks{E-mail:stefano.cavazzani@unipd.it}, V.Zitelli $^{2}$ \\
$^{1}$Department of Astronomy, University of Padova, Vicolo
dell'Osservatorio 3, I-35122, Padova, Italy\\
$^{2}$INAF-Osservatorio Astronomico di Bologna, via Ranzani 1, I-40127, Bologna, Italy\\
}

\begin{document}

\date{Accepted 2012 October 30.  Received 2012 October 19; in original form 2012 March 30.
}

\pagerange{\pageref{firstpage}--\pageref{lastpage}} \pubyear{2009}

\maketitle

\label{firstpage}

\begin{abstract}

In this paper we have evaluated the amount of available telescope time at four interesting sites for astronomical instrumentation. We use the {\bf{\it GOES 12}} data  for  the years 2008 and 2009. We use a homogeneous methodology presented in several previous papers to classify the nights as clear (completely cloud-free), mixed (partially cloud-covered), and covered. Additionally, for the clear nights, we have evaluated the amount of satellite stable nights which correspond to  the amount of ground based photometric nights, and the clear nights corresponding to the spectroscopic nights.
We have applied this model to two sites in the Northern Hemisphere (San Pedro Martir (SPM), Mexico; Iza$\tilde{n}$a, Canary Islands) and to two sites in the Southern Hemisphere (El Leoncito, Argentine; San Antonio de Los Cobres (SAC), Argentine). We have obtained, from the two years considered, a mean amount of cloud free nights of 68.6\% at Iza$\tilde{n}$a, 76.0\% at SPM, 70.6\% at Leoncito and 70.0\% at SAC. We have evaluated, among the cloud free nights, an amount of stable nights of 62.6\% at Iza$\tilde{n}$a, 69.6\% at SPM, 64.9\% at Leoncito, and 59.7\% at SAC.

\end{abstract}


\begin{figure*}
  \centering
  \includegraphics[width=16cm]{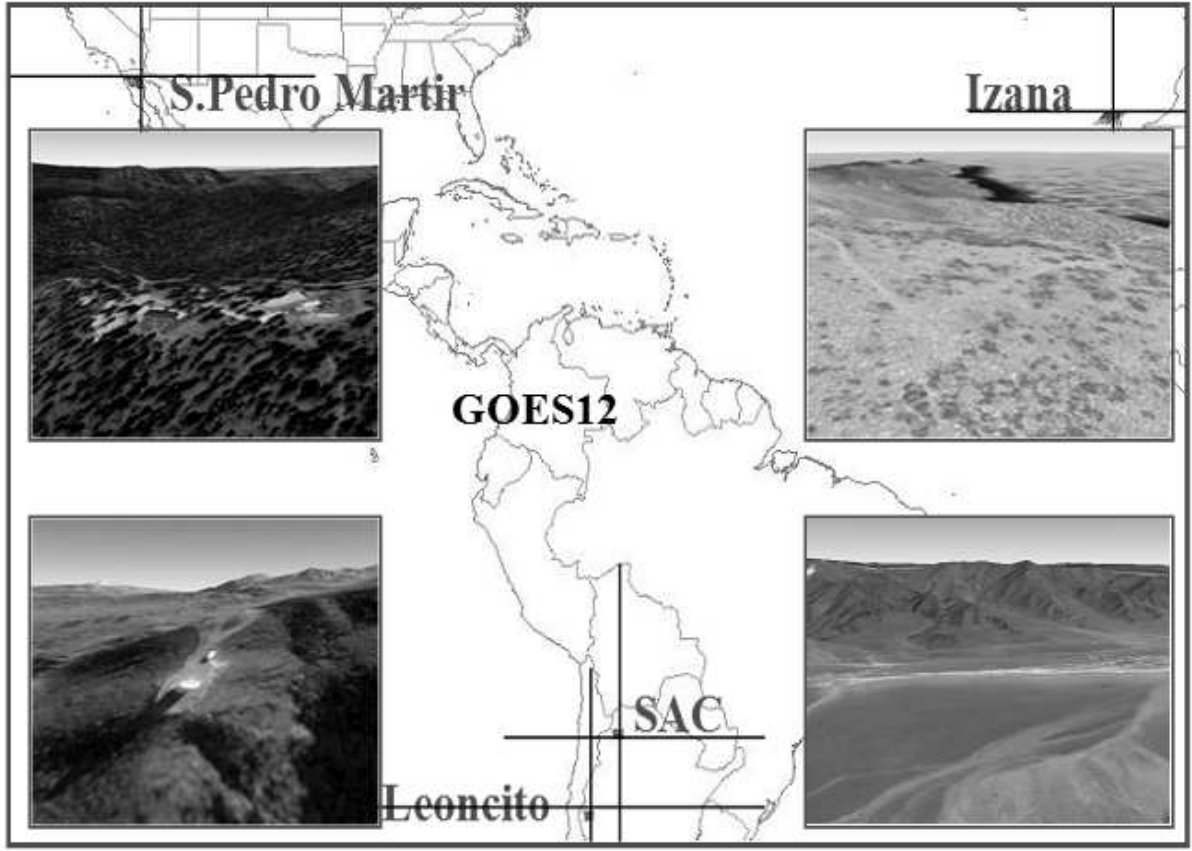}
  \caption{Location of the four sites involved in the analysis. As seen in the inserts (modified from Google Earth) the selected sites presents very different topographic
conditions. The position of {\bf{\it GOES 12}} satellite projected on the map (McIDAS-V Map).}
             \label{m}
\end{figure*}

\section{Introduction}

The basic requirement for astronomical sites is a good atmospheric transparency. Moreover, the duty cycle is mainly linked to the cloud coverage and atmospheric water vapour content.
For this reason in this paper we have concentrated our analysis on the evaluation of the amount of the nights suitable for observations. A complete survey of this parameter for each site under analysis is very difficult because standard criteria of classification do not exists. In the last century the evaluation of the night time clear fraction was based mainly on visual inspection of the sky, reported in the observational logbooks of the telescopes. This method is not only dependent on the experience of the observer, but it requires also a long time coverage. This approach is also time-consuming and expensive when applied to several sites, as is the case for the candidate sites of this analysis.
Satellite data are very useful and produced a great improvement in this field. The long time-series of data now available, from satellites in geostationary orbits, is particularly useful, as shown by Eramsus and Sarazin (\cite {era02}). Moreover such data allow to investigate simultaneously several sites, and give also short- and long-time trends. 
The satellite site testing analysis evolved with the use of multi-channel data, in particular at infrared wavelengths.
It allows to detect temperatures at different altitudes, discriminating parameters which produce the cloud coverage. 
In this study we extracted the night-time data to derive parameters to assess the amount of useful nights, using the Geostationary Operational Environmental Satellite 12 ({\bf{\it GOES 12}}) archive. {\bf{\it GOES 12}} data are analysed using a code which correlates several bands. We decided to use only GOES data because we have a uniform set of GOES data validated in different situations and at different sites. We know that Meteosat could be in a more favourable position with respect to the geographical location of Iza$\tilde{n}$a, but Meteosat data, available on the Web-page\footnote{http://www.eumetsat.int}, are being obtained already as daily averages. The night-time temporal resolution, which is crucial for the purpose of our analysis, is missing. The validation of our code was done using data of two important sites for the optical astronomy: El Roque de Los Muchahcos (ORM, Canary Islands), and Cerro Paranal (Chile), two very different sites with different climatic conditions. The results of the validation are published in Cavazzani at al. (\cite {cava11}), Della Valle et al. (\cite {dv10}).\\
We point out that our analysis is connected with the traditional definitions of the astronomical nights (clear, mixed and unusable). We are aware that there are interesting packages of data such as ECMWF\footnote{www.ecmwf.int} (Dee et al. (\cite{dee})), FriOWL\footnote{http://archive.eso.org/friowl-45/} ( Sarazin et al. (\cite{sara}), Graham et al. (\cite{graham}), (\cite{graham1})) and Giovanni \footnote{http://disc.sci.gsfc.nasa.gov/giovanni/overview/index.html} (Acker and Leptouk (\cite{giova07})), used for wide scale and long term climatic trends.
We intend to use them in a next work dedicated to long term cloud coverage. For the 
present paper these data do not have the required temporal and space resolution we need. In addition they have to be calibrated in terms of astronomical properties. Specific studies should be done on the sites.
In this paper we present the  evaluation of  the amount of usable nights for four interesting sites for astronomical instrumentation. The sites are located in  Argentina, Spain (Canary Islands) and Mexico. For each site we additionally report the  mean temperature and rainfall obtained from literature to have a more complete information of each analysed site. Table \ref{ST} shows the geographic positions of the sites, while Figure \ref{m} shows the location of the interested sites. The tiles of Figure \ref{m} show how different is the morphology of each site.  All the candidates sites are located in the subtropical belt, and  the Canary Islands site is the only one centred in a wide homogeneous oceanic air mass.\\

\section{Satellite Based Data}
\label{goes12}

In the last decades the site testing was greatly improved with the use of the satellite data and the traditional meteorological information (Eramsus and Sarazin (\cite {era02})). Varela et al. (\cite{varela}) give an exhaustive review  of the most useful satellites for site testing. 
Our analysis utilize information provided by the GOES satellites system, an American geosynchronous weather facility of the National Oceanic and Atmospheric Administration (NOAA).
GOES system is in a geosynchronous orbit allowing the satellites to stay continuously over one position on the surface. The geosynchronous orbit is about 35800 km above the Earth, high enough to allow the satellites a full-disc view of the Earth. In our analysis we use the data from {\bf{\it GOES 12}} satellite which is located at 75 deg W longitude over the equator. Full hemisphere images are simultaneously produced in several bands.
The mission is carried out by the primary instrument, the imager, which is a five channel instrument sensing the emissivity from the Earth's surface and atmosphere. We have selected the infrared thermal bands because we are interested in  the night time cloud coverage. Table \ref{BAND} shows the characteristics of the bands selected in this analysis, including the central pass-band and the spatial resolution computed at nadir of each site.\\
A detailed discussion about the advantages of this satellite is presented by
Cavazzani et al. (\cite{cava10}), Cavazzani et al. (\cite{cava11}) and by Della Valle et al. (\cite {dv10}). Each infrared band detects clouds at different heights. This is due to the Planck shape of its weighting function (WF), specific for each band, that presents a maximum of efficiency at different altitude according to the different selected band. The water vapour band (channel 3, band 3 hereafter B3, centered at $6.7~\mu m$ and sensitive between $6.5-7.0~\mu m$) is able to detect high altitude cirrus clouds, up to 8 km. The cloud coverage channel (channel 4, band 4 hereafter B4, centered at $10.7~\mu m$ and sensitive between $10.2-11.2~\mu m$) can detect middle level clouds, at about 4 km, and the $CO_2$ band (channel 6, band 6 hereafter B6, centered at $13.3~\mu m$)  senses  small particles such as fog, ash and low level clouds, at about 3 km\footnote{http://goes.gsfc.nasa.gov/}. 
The output of the detector is proportional to the energy reaching the sensor per unit time (radiance) that is equivalent to the brightness temperature. If clouds are not present during the night, the emission at B4 ($10.7~\mu m$) reaching the satellite is not absorbed  by the atmosphere so the measured radiance values are due to the emission from surface.

\begin{table}
 \centering
 \begin{minipage}{80mm}
  \caption{Geographic characteristics of the analyzed sites and{\bf{\it GOES 12}} satellite. The view angle is obtained through the formula $\theta=\sqrt{(\Delta LAT)^{2}+(\Delta LONG)^{2}}$.}
   \label{ST}
  \begin{tabular}{@{}lcccc@{}}
  \hline

  site      &LAT.&      LONG. & Altitude & View \\
            &    &            &  Km      &     Angle        \\
 \hline
 Leoncito    &   $-31^{\circ}47'$  &  $-69^{\circ}17'$  &  $2.552$   &  $32^{\circ}00'$  \\
 S.Pedro Martir   &   $31^{\circ}02'$  &  $-115^{\circ}29'$  &$2.800$     &  $51^{\circ}00'$    \\
 Iza$\tilde{n}$a   &   $+28^{\circ}18'$  &  $-16^{\circ}29'$  &  $2.373$   &  $61^{\circ}00'$   \\
 SAC    &$-24^{\circ}$ $02'$ &$-66^{\circ}$ $14'$    &  $3.600$ & $26^{\circ}00'$  \\
 \hline
{\bf{\it GOES 12}}     &   $+0^{\circ}00' $  &  $-75^{\circ}00'$  & $35800$    &                   \\
 \hline

\end{tabular}
\end{minipage}
\end{table}

\begin{table}
 \centering
 \begin{minipage}{80mm}
  \caption{{\bf{\it GOES 12}} bands and resolution at Nadir with their weighting functions (WF).}
   \label{BAND}
  \begin{tabular}{@{}lcccc@{}}
  \hline
                & window  & pass-band     & 1px resolution & WF     \\
                &         & $[\mu m]$     &    $[km]$     &  $[m]$     \\
 \hline
 \textit{BAND3} &$H_{2}O$   &   $6.50-7.00$ &  $4$   &$8000$\\
	\textit{BAND4}   & $IR$   & $10.20-11.20$ & $4$    &$4000$\\
	\textit{BAND6}&   $CO_{2}$  &    $13.30$ &  $8$       &$3000$\\
 \hline

\end{tabular}
\end{minipage}
\end{table}

\section{Satellite data acquisition}
\label{satacqui}

The main advantage of GOES with respect to non geostationary  satellites
 is that GOES has a large field of view. Figure \ref{m} shows the area of the field of view covering the sites interested in this analysis. Data are freely  available by the Comprehensive Large Array-data Stewardship System (CLASS), an electronic library of NOAA environmental data\footnote{www.class.ngdc.noaa.gov}, and are stored as rectified full earth disk images in a format called AREA files.
We processed data using \textit{McIDAS-V} Version 1.0 beta4, a free software package.
We have selected a region of the image for downloading, centered on each site.
For each hour and for each band we have processed data as following:

\begin{itemize}

	\item Extraction of $1^\circ \times 1^\circ$ sub-matrix centered on the coordinates of each site, at 20:45, 23:45, 02:45, 05:45, 8:45 local time for each night.
	\item Computing of the mean matrix value at 20:45, 23:45, 02:45, 05:45, 8:45 of each night and for each site.
	\item The total number of images used in this analysis is $N_{Tot}=43200$
	
\end{itemize}

\section{Resolution Correlation Matrix}

\begin{figure}
  \centering
  \includegraphics[width=8.5cm]{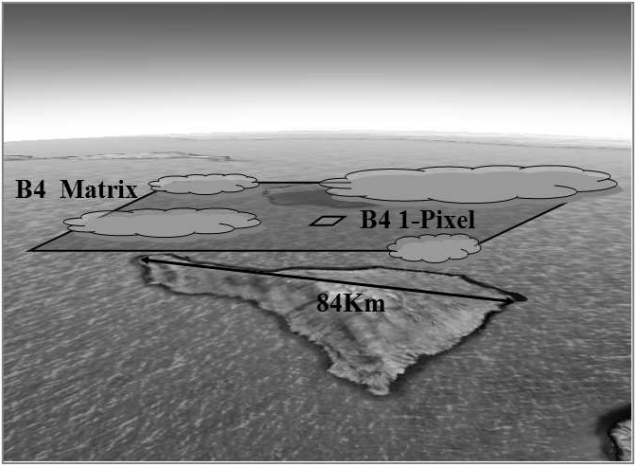}
  \caption{Single pixel (small rectangle $4Km\times 4Km$) vs. matrix (wide rectangle $100Km\times 100Km$) coverage comparison at Iza$\tilde{n}$a. The length of the island is approximately $84km$. Figure shows that the use of the single pixel is not sufficient in the case of a partially covered sky to define the properties of the night.}
             \label{ima1}
\end{figure}

\begin{figure}
  \centering
  \includegraphics[width=8.5cm]{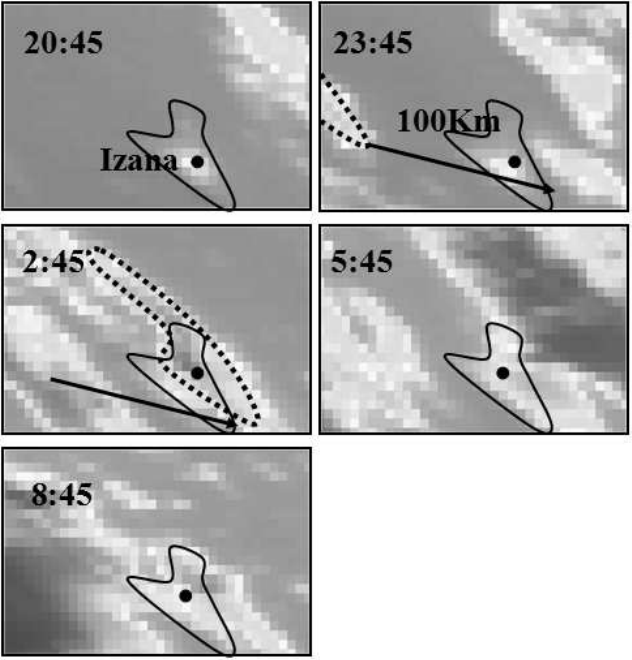}
  \caption{Clouds time evolution of a partially covered night (the dotted line represents a moving cloud) at Iza$\tilde{n}$a.}
             \label{ima2}
\end{figure}

In this analysis we used for each site a $1^{\circ}\times 1^{\circ}$ matrix, which corresponds to a linear projection of about 100 km. We have shown in Cavazzani et al. (\cite{cava10}) that the trend of the mean matrix value is closely correlated with the trend of the single pixel value, with a correlation coefficient of $>95\%$ for satellite angles of view $<60^{\circ}$. The use of the matrix mean value drastically reduces the noise of the images.
Figure \ref{ima1} shows that a single pixel (in the small rectangle), $4Km\times 4Km$ wide, does not cover a sufficiently wide sky area above the site because it corresponds to a small geometric field of view as seen from the ground.
In fact, the B4 weighting function, obtained using the calibration page\footnote{ http://cimss.ssec.wisc.edu/}, observes an atmospheric layer at an altitude of about $4000m$ (see Table \ref{BAND}). We emphasize that the B4 band is not able to detect clouds below this altitude, because its  weighting function reaches its maximum efficiency at 4 Km, and drops just below this altitude. Clouds below this limit do not directly affect the measurements in this band. This implies that our statistics match more closely on upper limit. The use of the matrix allows to observe the entire sky above the site.
In addition, if we have clouds in motion, as in  the case of partial cloud coverage, the use of the matrix allows us to observe them.\\
Furthermore, the use of the single pixel, does not show the temporal cloud evolution giving instantaneous clear time instead of mixed or cloudy time.
Figure \ref{ima2} shows the evolution of a partially covered night at Iza$\tilde{n}$a taken on February 7 2008.\\
We see five images of the island in the B4 band taken at 20:45-23:45-2:45-5:45-8:45 (Local Time). The frame of 20:45 shows Iza$\tilde{n}$a completely cloud free. The second image shows a thick cloud (dashed line) that enters in the matrix field of view moving in the direction of the arrow. The frame taken at 2:45 shows the movement of this cloud.
We have estimated the displacement of the cloud to be about 100 km. It comes on top of the analysed site in the next  three hours.
With a simple calculation we  estimated the velocity of this cloud of about $30 km/h$.\\
In conclusion, the use of the matrix combined with the analysis of 5 images for each night gives us a good time resolution. 
The matrix allows an intrinsic increase of the spatial-time resolution: the matrix observes the temporal evolution of the entire sky above the site.
Such temporal resolution and coverage can not, therefore, be reached by the models based on analysis of single pixel and the single image for each night.

\section {Satellite night classification}

\subsection{Clear, Mixed, Covered Nights}
\label{cmcnc}

The satellite night classification is based on  the B4 band flux variability in each month.
We assume that the maximum brightness temperature $T^{Max}_{B}$ in the B4 band occurs if the night is clear. This occurs because, when we have an optically thick cloud in the path, the brightness temperature detected by the satellite drops according to the top temperature of the cloud. The model is accurately described in Cavazzani et al. (\cite{cava10}) and Cavazzani et al. (\cite{cava11}). We search  the maximum brightness temperature $T^{Max}_{B}$ for each month as the  reference temperature.
Moreover we compute for each month a threshold as the statistical mode, night by night, of just the differences of the brightness temperature at 23:45 and 8:45 (we have estimated that at these times we have the maximum and minimum values of the brightness temperature). 
We define  $1\sigma$ value through the formula:

\[Mode~Night~Satellite~Temperature=1\sigma =T^{23:45}_{B}- T^{8:45}_{B}\]         

The nights are then classified according
to the value of $T_{B}$:

\begin{itemize}

	\item $T^{Max}_{B}-T_{B}\leq2\sigma\Longrightarrow$ Clear
	\item $2\sigma<T^{Max}_{B}-T_{B}\leq3\sigma\Longrightarrow$ Mixed
	\item $T^{Max}_{B}-T_{B}>3\sigma\Longrightarrow$ Covered
	
\end{itemize}

where $T_{B}\Rightarrow$ is the computed mean brightness temperature of the $1^\circ\times1^\circ$ matrix at 20:45, 23:45, 02:45, 05:45, 8:45 for each night and in the B4 band.
All the above described procedures were validated using the Paranal and La Palma observing logbooks. Table 4  of Cavazzani et al. (\cite{cava11}) shows the results of the validation.

\subsection{Clear, Stable Nights }
\label{csnc}

To have a more reliable comparison between satellite and ground based classification we introduce the concept of a stable night. This concept takes into account all the selected bands defining an atmospheric correlation function $F_{C.A.}(t)$   in the following way 

\begin{equation}
F_{C.A.}=I_{\lambda_{3}}-[I_{\lambda_{6}}-I_{\lambda_{4}}]
\end{equation}

Where $I_{\lambda_{3}}$, $I_{\lambda_{4}}$ and $I_{\lambda_{6}}$ are the monochromatic emission fluxes from the Earth surface in each band as received by the sensor.
$F_{C.A.}(t)$ is a value computed for each detected hour and for each night. We would say that this model takes into account auto-corrections of atmosphere: for instance if two high layers have a positive oscillation and one lower layers has an equal magnitude oscillation, but negative, the $F_{C.A.}(t)$ remains constant. From the physical point of view this means that there are no relative changes in the atmospheric properties at the two layers, so they are affected by uniform air masses.\\
Considering the best fit of the monthly plot of $F_{C.A.}(t)$ we obtain the following classification of the nights:
\begin{itemize}

	\item $\left|T_{B}-T^{Trendline}_{B}\right|\leq\left|1\sigma\right|\Longrightarrow$ Stable
	
\end{itemize}

where:

\begin{enumerate}

  \item $T^{Trendline}_{B}\Rightarrow$  brightness temperature obtained from the fit of the $F_{C.A.}(t)$ at the given hour
	\item $T_{B}\Rightarrow$ $F_{C.A.}(t)$ current brightness temperature of the $1^\circ\times1^\circ$ matrix
\end{enumerate}

With these definitions we obtain the fraction of clear and stable nights of the analysed sites as shown and discussed in the following sections.

\section{Sites analysis}

In this section we present the classification of the nights at 
each site using  the code described above. In order to have a more general information. The results are compared with literature data.

\subsection{Sites located in the North Hemisphere}

 \subsubsection{Iza$\tilde{n}$a (Canary Island) }

The climate of all the Canary Islands, located in north of the Tropic of Cancer, is modulated by  the Azores Anticyclone.  Iza$\tilde{n}$a Observatory (IZO) has an Atmospheric Research Center (IARC) in the mountains of the   island of Tenerife.  The climatic characteristics of IZO are driven by the altitude. IZO is located at 2400 meters above sea level, higher than the quasi-permanent temperature inversion layer associated to the trade-winds regime. The inversion layer separates the moist marine boundary layer from the dry, free troposphere. 
Table \ref{IZOdata} shows the monthly distribution of temperature and rain obtained using the Agencia Estatal de Meteorología (AEMET) database to show the seasonal trend. The average of temperatures obtained on this time baseline is $9.2^{\circ}$C. In 2008 the mean annual temperature was $9.8^{\circ}$C, while in 2009 it was $10.2^{\circ}$C. These two values are well above the mean values in Table \ref{IZOdata}. Moreover we see that the precipitations occur mainly in winter time when Atlantic low pressure systems pass over the Canary Islands.

  \begin{table}
     \centering
     \begin{minipage}{80mm}
      \caption{Mean monthly temperature (1920-2010),rain(1971-2000) and wind velocity at Iza$\tilde{n}$a (www.Iza$\tilde{n}$a.org). }
       \label{IZOdata}
       \begin{tabular}{@{}lcc@{}}
      \hline
    month        & Mean temp.  & Precipitation     \\
                &   (deg C)    & (mm)             \\
     \hline
     January     & 3.8  & 86.6  \\
     February    & 3.9  & 64.3  \\
     March       & 5.3  & 65.0   \\
    April        & 6.8  & 25.6  \\
     May         & 9.3  & 13.7  \\
     June        & 13.2 & 0.5  \\
     July        & 17.4 & 0.4  \\
     August      & 17.4 & 2.4  \\
     September   & 13.6 & 14.8  \\
    October      & 9.7  & 36.0  \\
     November    & 6.6  & 50.0  \\
     December    & 4.2  & 73.4   \\
     \hline    
    \end{tabular}
    \end{minipage}
    \end{table}

The first evaluation of the available observing time at Tenerife, is by Murdin (\cite{murdin85}) with 61\%  of photometric hours   and 14\% of spectroscopic hours, computed from February to September 1975. 
 
The results obtained by  our code using the {\bf{\it GOES 12}} satellite data are shown in Table \ref{izonight}, where we present the percentage of covered, clear and stable nights for 2008 and 2009. The number of clear nights
shows a large variability over the two years, including the summer time, which is a typically dry period. Taking into account the same period from  Murdin, February-September, we obtain from satellite 79\% of clear nights for  2008 and 69\% for 2009, considerably higher values with respect to those of Murdin. Can we ascribe these differences to a yearly variability or to different evaluation criteria? 
 
  \begin{table}
     \centering
     \begin{minipage}{200mm}
      \caption{{\bf{\it GOES 12}} night time classification at Iza$\tilde{n}$a . }
       \label{izonight}
        \resizebox*{0.45\textwidth}{!} {
      \begin{tabular}{@{}lcccccccc@{}}
      \hline
                &\multicolumn{4}{c}{Iza$\tilde{n}$a 2008}&\multicolumn{4}{c}{Iza$\tilde{n}$a 2009}  \\
                 \hline
     Month       & Covered &  Mixed & Clear & Stable & Covered &  Mixed  & Clear & Stable  \\
     \hline
     January     & 39.9  & 6.3  & 53.8 & 48.3 & 34.2 & 14.2 &51.6 & 50.8  \\
     February    & 21.8  & 5.6  & 72.6 & 62.1 & 30.6 & 14.5 &54.9 & 49.2  \\
     March       & 33.8  & 9.0  & 57.2 & 47.6 & 27.8 & 11.1 &61.1 & 54.9  \\
    April        & 9.3   & 2.9  & 87.9 & 79.3 & 10.7 & 12.9 &76.4 & 66.4  \\
     May         & 5.6   & 7.0  & 87.4 & 74.1 & 9.1  & 18.2 &72.7 & 68.5  \\
     June        & 0.0   & 4.9  & 95.1 & 93.1 & 16.7 & 7.6  &75.7 & 66.7  \\
     July        & 3.4   & 4.1  & 92.5 & 78.6 & 11.0 & 9.0  &80.0 & 76.6  \\
     August      & 10.9  & 7.0  & 82.1 & 81.4 & 8.6  & 8.6  &82.8 & 72.9  \\
     September   & 30.9  & 9.2  & 59.9 & 59.2 & 34.9 & 16.6 &48.5 & 45.0  \\
     October     & 9.2   & 6.9  & 83.9 & 78.2 & 20.1 & 10.9 &69.0 & 67.2  \\
     November    & 27.3  & 10.5 & 62.2 & 53.8 & 20.3 & 16.1 &63.6 & 59.4  \\
     December    & 42.6  & 11.5 & 45.9 & 41.0 & 51.1 & 19.9 &29.0 & 27.7 \\
     \hline
  Mean           & 19.6  &7.1   & 73.4 & 66.4 & 22.9 & 13.3 & 63.8 & 58.8 \\

    \end{tabular}
    }
    \end{minipage}
    \end{table}

 \subsubsection{S.Pedro Martir, Baja California }
 
The second site analysed is the observatory at  San Pedro Martir in  Baja California, Mexico. 

Tapia et al. \cite{tapia07} give the results of more than three decades of site characterizations. This long statistical analysis of cloud coverage shows that the fraction of nights lost due to bad weather is about 21.4\% in the period July 1982 to December 2006. From January 1984 to December 2006, 64.1\% of the nights were of photometric quality and 80.3\% were of spectroscopic quality. The average relative humidity near the ground was 54\% (Tapia (\cite{tapia92})) with a large seasonal dependence. Spring and autumn are the best seasons in terms of cloudless and low humidity nights, while winter is affected by the tails of North Pacific storms and mid summer is characterized by a mild monsoon season. The data are based on the $2.1m$ telescope observing log. For the same site, Erasmus and van Rooyen (\cite{era06}) found an amount of useful time of 81.6\%, obtained using the IR GOES satellite bands computed for the period June 1997 to May 1998, showing, for the same period, a good agreement with the Tapia data (about $\pm5\%$).
Using the weather station installed at this site, we collected temperature, wind speed and rainfall data for the years 2007 to 2010 using the SPM weather Web-page\footnote{http://www.astrossp.unam.mx/weather15/}.
Table \ref{SPMmeteo} summarizes the monthly distribution of these parameters.

  \begin{table}
     \centering
     \begin{minipage}{80mm}
      \caption{Mean monthly temperature,rain and wind velocity at SPM of the years 2007 to 2010. 
      (http://www.astrossp.unam.mx/weather15/)  }
       \label{SPMmeteo}
      \begin{tabular}{@{}lccc@{}}
      \hline
    
     month        & Mean temp.  & Precipitation  & wind speed   \\
                 &   (deg C)    & (mm)      & (m/s)         \\
     \hline
     January     & 1.2  & 18.5 & 4.1\\
     February    & 1.3  & 1.2  & 3.6\\
     March       & 4.1  & 3.6  & 3.3 \\
    April        & 5.5  & 2.9  & 3.7\\
     May         & 8.8  & 1.5  & 3.4\\
     June        & 13.9 & 3.2  & 3.7\\
     July        & 15.7 & 62.2 & 2.3\\
     August      & 15.2 & 80.9 & 2.9\\
     September   & 12.3 & 9.9  & 3.5\\
     October     & 8.7  & 14.2 & 4.7\\
     November    & 5.9  & 11.8 & 4.2\\
     December    & 3.8  & 3.4  & 3.8 \\
     \hline
    
    \end{tabular}
    \end{minipage}
    \end{table}

  \begin{table}
     \centering
     \begin{minipage}{200mm}
      \caption{{\bf{\it GOES 12}} night time classification at SPM. }
       \label{SPMnight}
        \resizebox*{0.45\textwidth}{!} {
      \begin{tabular}{@{}lcccccccc@{}}
      \hline
                &\multicolumn{4}{c}{S.Pedro Martir 2008}&\multicolumn{4}{c}{S.Pedro Martir 2009}  \\
                 \hline
     Month       & Covered &  Mixed & Clear & Stable & Covered &  Mixed  & Clear & Stable  \\
     \hline
     January      & 32.2 & 2.8 &65.0 & 58.7& 18.3 & 10.0 &71.7 & 69.2   \\
     February     & 13.7 & 8.1 &78.2 & 75.8& 32.3 & 9.7 &58.0 & 54.8    \\
     March        & 4.1  & 2.1 &93.8 & 84.1& 16.7  & 7.6 &75.7 & 63.2   \\

    April         & 3.6  & 0.7 &95.7 & 84.3& 7.1  & 7.1 &85.8 & 80.7    \\
     May          & 8.4  & 4.2 &87.4 & 83.2& 7.0  & 4.2 &88.8 & 71.3    \\
     June         & 0.0  & 1.4 &98.6 & 97.2& 22.9  & 14.6 &62.5 & 56.3  \\
     July         & 14.5 & 6.9 &78.6 & 75.2& 22.1 & 7.6 &70.3 & 60.0    \\
     August       & 16.3 &17.1 &66.6 & 58.9& 8.6 & 2.1 & 89.3 & 82.9    \\
     September    & 9.2  & 5.9 &84.9 & 80.3& 13.0  & 7.1 & 79.9 & 74.0  \\
     October      & 0.0  & 3.4 &96.6 & 89.7& 11.5  & 9.2 & 79.3 & 70.1  \\
     November     & 30.1 &11.2 &58.7 & 54.5& 15.4 & 4.2 & 80.4 & 72.7   \\
     December     & 42.6 &11.5 &45.9 & 42.6& 46.6 &22.7 & 30.7 & 29.8  \\
     \hline
  Mean            & 14.6 & 6.3 & 79.2 & 73.7& 18.5 & 8.8 & 72.7 & 65.4  \\

    \end{tabular}
    }
    \end{minipage}
    \end{table}
    
Table \ref{SPMnight} shows the  classification of the nights obtained using the {\bf{\it GOES 12}} satellite. For this site we also see a large variability for the two years. The percentages of the nights are not very different from those obtained using the ground based data. The satellite gives an additional 10\% of clear nights.

  \subsection {Sites located at the South Hemisphere}
 
  \subsubsection{S. Antonio de los Cobres, Argentina }
   
S. Antonio de los Cobres  is located at 3600 m above sea level, and is located in the province of Salta, 164 km away from Salta. Some limited meteorological data are available for the period 2001-2010, and the monthly values can be found in the official report of the two sites in Argentina compiled by the Argentinian and Brazilian parties (version 1.1, July 6 2011). On the basis of the temperatures acquired at the meteorological station of the {\it Gendarmeria National}, which gives the monthly averages over ten years (2001-2010), we found that the maximum temperature occurs in November (14.4 $^0$C) while  the minimum temperature occurs in August (-7 $^0$C).
From the {\bf{\it GOES 12}} satellite we obtain the percentages of available time  presented in Table \ref{SACnight}. These results are almost in agreement with a similar analysis of Erasmus and Maartens (\cite{era2001}) using the {\bf{\it GOES 8}} satellite. They derive  percentages of 75\% of clear time and 5\% of partially cloudy time at the coordinates of Macon (Lat -24 37' Long -67 19'), for the period 1993-1999.

   \begin{table}
         \centering
         \begin{minipage}{200mm}
          \caption{{\bf{\it GOES 12}} night time classification at SAC. }
           \label{SACnight}
            \resizebox*{0.45\textwidth}{!} {
          \begin{tabular}{@{}lcccccccc@{}}
          \hline
                    &\multicolumn{4}{c}{S.Antonio de Los Cobres 2008}&\multicolumn{4}{c}{S.Antonio de Los Cobres 2009}  \\
                     \hline
         Month       & Covered &  Mixed & Clear & Stable & Covered &  Mixed  & Clear & Stable  \\
         \hline
         January     &67.8 &13.3 & 18.9 & 16.8 & 25.8  & 24.2 &50.0 & 47.5   \\
         February    &15.3 &16.9 & 67.8 & 56.5 & 28.2  & 14.5 &57.3 & 54.8   \\
         March       &10.3 &15.2 & 74.5 & 66.2 & 27.8  & 20.8 &51.4 & 50.7  \\
        April        &6.4  &9.3  & 84.3 & 72.9 & 7.1   & 9.3  &83.6 & 72.1   \\
         May         &7.0  &4.2  & 88.8 & 76.9 & 4.2   & 2.1  &93.7 & 72.0   \\
         June        &13.9 &9.7  & 76.4 & 67.4 & 13.9  & 6.9  &79.2 & 63.9   \\
         July        &0.0  &2.1  & 97.9 & 74.1 & 26.9  & 19.3 &53.8 & 49.7   \\
         August      &3.1  &3.1  & 93.8 & 76.0 & 7.9   & 8.6  & 83.5 & 65.7   \\
         September   &6.6  &13.8 & 79.6 & 64.5 & 8.3   & 10.1 & 81.6 & 65.1  \\
         October     &39.7 &5.7  & 54.6 & 49.4 & 20.7  & 10.9 & 68.4 & 60.3 \\
         November    &4.2  &14.0 & 81.8 & 72.0 & 12.6  & 21.7 & 65.7 & 49.7  \\
         December    &29.5 &21.3 & 49.2 & 47.5 & 42.6  & 12.1 & 45.3 & 41.1 \\
         \hline
         Mean        & 17.4 & 10.7 & 72.3 & 61.7& 18.8 & 13.4 & 67.8 & 57.7 \\

        \end{tabular}
        }
        \end{minipage}
        \end{table}

\subsubsection{Satellite Wind Analysis}
\label{swa}

\begin{figure}
  \centering
  \includegraphics[width=8.5cm]{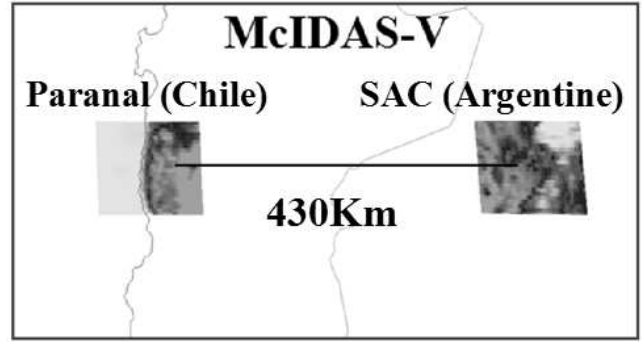}
  \caption{Paranal and SAC matrix (McIDAS-V Map). 
  The geographical proximity allows the same calibration of the empirical model.}
             \label{map2}
\end{figure}

\begin{figure}
  \centering
  \includegraphics[width=8.5cm]{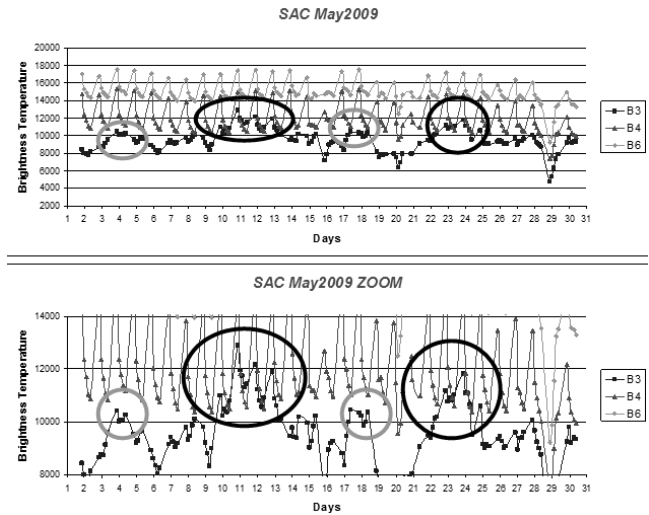}
  \caption{Analysis of the brightness temperature May 2009 at SAC. The gray ovals show perturbed points to be interpreted as windy episodes. Black ovals contains points possibly corresponding to the extremely strong wind.}
             \label{wind1}
\end{figure}

Analyzing the SAC site we noticed strong fluctuations of the B3.
In particular, we noticed that sometimes the B3 value exceeded the B4 value (see Figure \ref{wind1}).
These episodes occur only in rare cases at Paranal, although to a lesser extent. From the Paranal Web-site we  noticed that a strong ground wind corresponds to these events.
For this reason we started a  preliminary study from the satellite wind analysis, using the B3 and B4 bands, for  this site which we discuss in this Section. We plan to improve this model in the future using  more data to detect the presence of different wind phenomena affecting the results.
In this paper we present two different and preliminary classifications:
a statistical classification based only on the trend of B3 and B4 and a physical classification based on site characteristics (altitude, temperature and satellite angle of view).
The physical classification gives us the opportunity to obtain an estimate of the wind speed. The  wind model is empirically  calibrated using the data of Paranal obtained from the Web-site. Figure \ref{map2} shows that the two sites lie at a  distance of about $430km$.

\subsubsection{Statistical Wind Classification}
\label{w}

Using the classification of  the ESO Web-page\footnote{http://archive.eso.org/asm/ambient-server} for Paranal we can do an initial classification of the wind speed:

\begin{itemize}
	\item Weak wind $\Rightarrow v < 10m/s$
	
	\[I_{\lambda_{4}}-I_{\lambda_{3}}>1\sigma
\]
	\item Strong wind $\Rightarrow 10m/s < v < 15m/s$
	
	\[0<I_{\lambda_{4}}-I_{\lambda_{3}}<1\sigma
\]

	\item Extremely strong wind $\Rightarrow v > 15m/s$

\[I_{\lambda_{4}}-I_{\lambda_{3}}<0
\]

\end{itemize}

\begin{figure*}
  \centering
  \includegraphics[width=16cm]{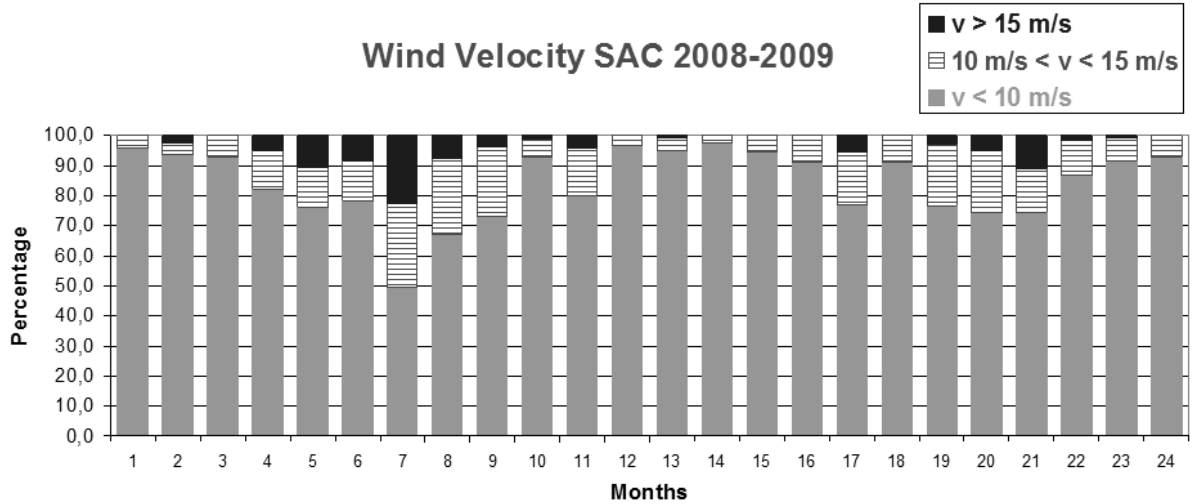}
  \caption{Estimated distribution of the wind speed at SAC (2008-2009) from satellite analysis. The gray bars show the monthly percentage of time with little or no wind ($v < 10m/s$), the dashed gray bars indicate the monthly percentage of time with strong winds ($10m/s < v < 15m/s$) and the black bars indicate the monthly percentage of time with extremely strong wind ($v > 15m/s$).}
             \label{wind}
\end{figure*}

\begin{figure}
  \centering
  \includegraphics[width=8.5cm]{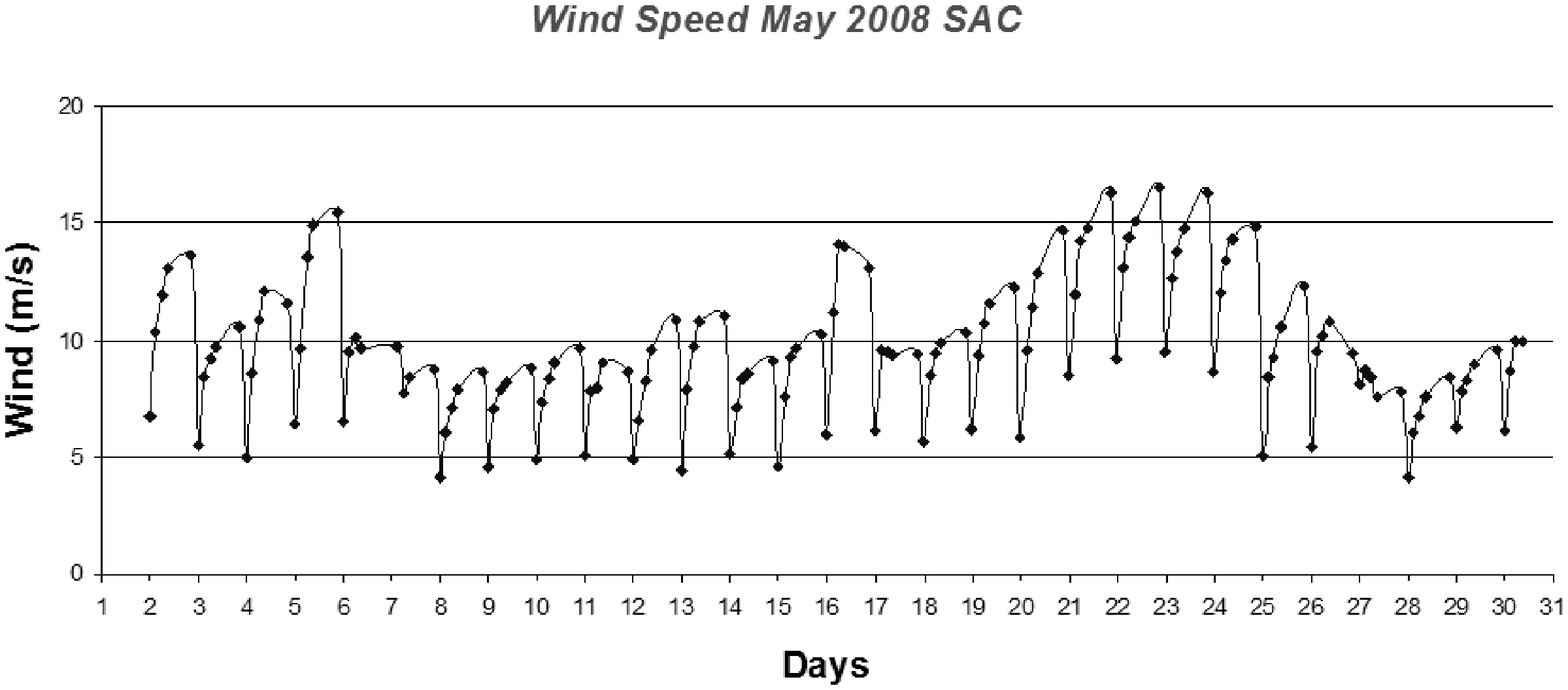}
  \caption{Figure shows the monthly trend of the wind speed $\left[\frac{m}{s}\right]$ obtained with the physical wind classification (May 2008 at SAC).}
             \label{wind2}
\end{figure}

Tables \ref{w1} and \ref{w2} show the results of this classification applied at SAC in  2008 and in 2009 respectively.
Figure \ref{wind} shows the two years histogram of this satellite analysis. July 2008 has been a very windy month followed by May 2008 as shown in Figure \ref{wind2} that plots the wind speed for the complete month of May 2008 at SAC.

\begin{table}
         \centering
         \begin{minipage}{80mm}
          \caption{{\bf{\it GOES 12}} velocity of wind classification at S. Antonio de Los Cobres (SAC) 2008.}
           \label{w1}
          \begin{tabular}{@{}lcccccccc@{}}
          \hline
                 
         Month        & Weak Wind & Strong Wind & E-Strong Wind \\
         \hline
         January     &95.8 &4.2 & 0.0   \\
         February    &93.6 &4.0 & 2.4  \\
         March       &93.1 &6.9 & 0.0 \\
         April        &82.1  &12.9  & 5.0   \\
         May         &76.2 &13.3  & 10.5  \\
         June        &78.5 &13.2  & 8.3   \\
         July        &49.4  &27.8  & 22.8   \\
         August      &67.4  &24.8 & 7.8  \\
         September   &73.1 &23.0& 3.9  \\
         October     &93.2 &5.7  & 1.1 \\
         November    &79.7  &16.1 &4.2  \\
         December    &96.7 &3.3 & 0.0  \\
         \hline
         Mean        & 81.6 & 12.9 & 5.5 \\

        \end{tabular}
        \end{minipage}
        \end{table}

\begin{table}
         \centering
         \begin{minipage}{80mm}
          \caption{{\bf{\it GOES 12}} velocity of wind classification at S. Antonio de Los Cobres (SAC) 2009.}
           \label{w2}
          \begin{tabular}{@{}lcccccccc@{}}
          \hline
                 
         Month        & Weak Wind & Strong Wind & E-Strong Wind \\
         \hline
         January     &95.0 &4.2 & 0.8   \\
         February    &97.6 &2.4 & 0.0  \\
         March       &94.4 &5.6 & 0.0 \\
         April        &91.4  &8.6  & 0.0   \\
         May         &76.9 &17.5  & 5.6  \\
         June        &91.0 &9.0  & 0.0   \\
         July        &76.5  &20.7  & 2.8   \\
         August      &74.3  &20.7 & 5.0  \\
         September   &74.5 &14.8 & 10.7  \\
         October     &86.8 &11.5  & 1.7 \\
         November    &91.6  &7.7 &0.7  \\
         December    &92.9 &7.1 & 0.0 \\
         \hline
         Mean        & 86.9 & 10.8 & 2.3 \\

        \end{tabular}
        \end{minipage}
        \end{table}

\subsubsection{Physical Wind Classification}
\label{ww1}

We can also to evaluate empirically the wind speed that we present in this Section.
Also in this case the model is calibrated using the down-loaded data from the Web-page of Paranal.
The wind speed is calculated as a function of the site altitude ($h(m)$) and the satellite angle of view ($\theta$).
The model uses the exponential trend of the Kolmogorov theory of turbulence combined with the kinetic theory of gases, and the wind velocity (expressed in $\left[\frac{m}{s}\right]$) is obtained through the formula:

\begin{equation}
	v = \sqrt{e^{\left[\frac{A}{B/cos\theta}\right]}}
\label{wv}
\end{equation}

where:

	\[A=\frac{I_{\lambda_{4}}-I_{\lambda_{3}}}{cos\theta}+\left\{\frac{\overline{\left|I_{\lambda_{4}}-I_{\lambda_{3}}\right|}}{h(m)}\right\}\cdot \frac{\overline{I_{\lambda_{3}}}}{cos\theta}+\frac{2h(m)}{cos\theta}
\]

and

	\[B=\overline{\left|I_{\lambda_{4}}-I_{\lambda_{3}}\right|}
\]

We emphasize that the model was empirically calibrated.
We note, however, that the equation \ref{wv} is based only on the altitude of the site and on the satellite data.
This allows us to estimate the wind speed without the need of ground data.

 \subsubsection{El Leoncito, Argentina }

This site is located in the San Juan Province. It is near Casleo Observatory, and close to the El Leoncito National Park.
The Observatory has its own meteorological station. The Web-page\footnote{www.casleo.gov.ar/weather/leonci\_weather.htm} shows the current data.
Rovero et al. (\cite{rovero}) give the monthly mean values of the last 10 years derived from the local meteorological station. Some of the most important meteorological data are given in Table \ref{w5}. 
Table \ref{leonight} shows the  percentages of available time from {\bf{\it GOES 12}} obtained for this site. We obtained an amount of about 71\% of clear time, in good agreement with the percentage of  
74\% of observing nights obtained by Rovero et al. (\cite{rovero}) from the analysis of more than 20 years of observing nights.

   \begin{table}
            \centering
            \begin{minipage}{80mm}
             \caption{Mean monthly meteorological parameters}
              \label{w5}
             \begin{tabular}{@{}lll@{}}
             \hline

            Parameters & Values  \\
            \hline
            Mean temperature &  5.5 $^0$C.(July) and 17.5 $^0$C. 
(January)   \\
            Maximum temperature & 20.3$^0$C.(June) and 30.5 $^0$C. 
(January)  \\
            Rel.hum. & 27-43 \% along the year \\
            Rainfall & 11-18 mm (December-March) \\
            & and 2-7 mm (April-November) \\
            Wind speed & 4.5-8.4 km/hr, mainly from SW (55 \%)\\
            & and from SE (30 \%)\\
            Maximum wind speed & 62-88 km/hr along the year   \\
            Hail & 1-2 days/yr   \\
            Frost & none  \\
            Lightning & 6-8 days/yr\\
            Maximum snowfall & 30 cm  \\
                                               \\

           \end{tabular}
           \end{minipage}
           \end{table}

     \begin{table}
           \centering
           \begin{minipage}{200mm}
            \caption{{\bf{\it GOES 12}} nights classification at Leoncito. }
             \label{leonight}
              \resizebox*{0.45\textwidth}{!} {
            \begin{tabular}{@{}lcccccccc@{}}
            \hline
                      &\multicolumn{4}{c}{Leoncito 2008}&\multicolumn{4}{c}{Leoncito 2009}  \\
                       \hline
           Month       & Covered &  Mixed & Clear & Stable & Covered &  Mixed  & Clear & Stable  \\
           \hline
           January     & 9.1   & 19.6 & 71.3 & 66.4  & 15.8  & 20.8  & 63.4 & 60.   \\
           February    & 27.4  & 9.7  & 62.9 & 62.1  & 13.7  & 6.5   & 79.8 & 67.7  \\
           March       & 17.9  & 2.8  & 79.3 & 73.8  & 13.2  & 12.5  & 74.3 & 72.9  \\
          April        & 17.1  & 8.6  & 74.3 & 72.1  & 1.4   & 4.3   & 94.3 & 83.6  \\
           May         & 15.4  & 7.0  & 77.6 & 69.2  & 38.5  & 23.8  & 37.7 & 34.3  \\
           June        & 15.3  & 6.9  & 77.8 & 72.2  & 27.8  & 7.6   & 64.6 & 56.3  \\
           July        & 37.0  & 5.9  & 57.1 & 55.6  & 13.8  & 13.1  & 73.1 & 62.8  \\
           August      & 21.7  & 12.4 & 65.9 & 58.9  & 27.1  & 11.4  & 61.5 & 58.6  \\
           September   & 35.5  & 7.9  & 56.6 & 54.6  & 20.1  & 14.8  & 65.1 & 58.6  \\
           October     & 9.2   & 6.3  & 84.5 & 77.6  & 33.9  & 9.2   & 56.9 & 50.0   \\
           November    & 5.6   & 7.0  & 87.4 & 86.7  & 16.8  & 6.3   & 76.9 & 65.0  \\
           December    & 9.8   & 9.8  & 80.4 & 78.7  & 21.3  & 7.1   & 71.6 & 59.6 \\
           \hline
           Mean        & 18.4  & 8.7  & 72.9 & 69.0 & 20.3 & 11.5 & 68.3 & 60.8 \\

          \end{tabular}
          }
          \end{minipage}
          \end{table}

\section{Discussion}

\begin{figure}
  \centering
  \includegraphics[width=8.5cm]{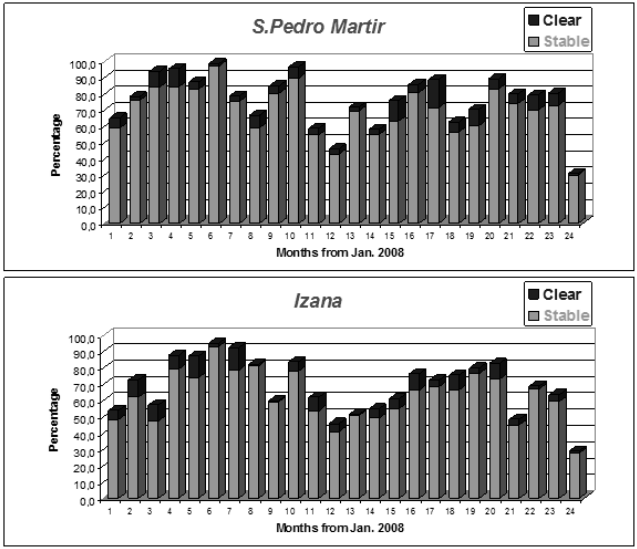}
  \caption{Distributions of stable and clear nights at S. Pedro Martir and Iza$\tilde{n}$a (2008-2009) obtained from {\bf{\it GOES 12}} satellite.}
             \label{stable}
\end{figure}

\begin{figure}
  \centering
  \includegraphics[width=8.5cm]{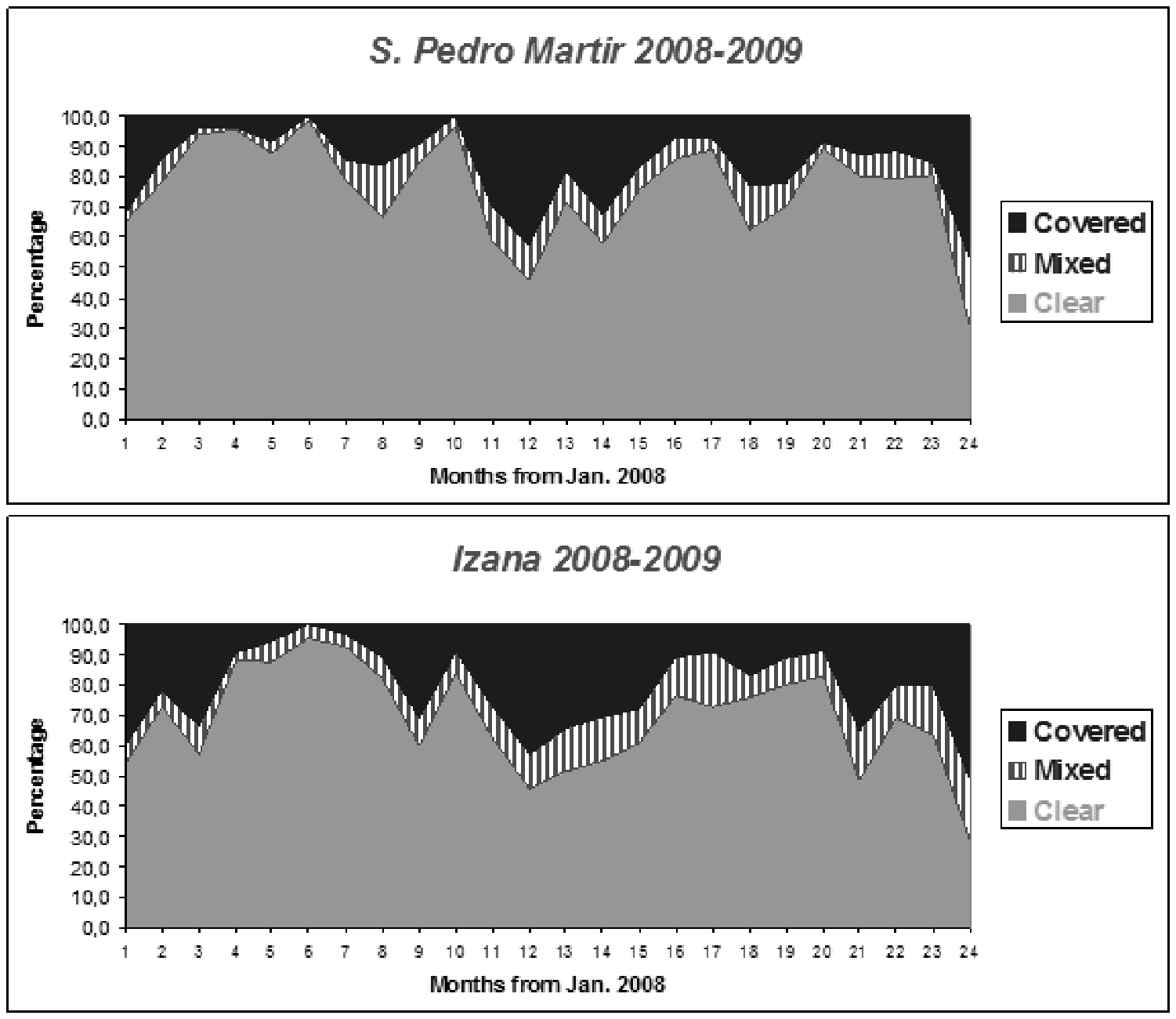}
  \caption{Distributions of clear, mixed and covered nights at S. Pedro Martir and Iza$\tilde{n}$a (2008-2009) obtained from {\bf{\it GOES 12}} satellite.}
             \label{clear}
\end{figure}

\begin{figure}
  \centering
  \includegraphics[width=8.5cm]{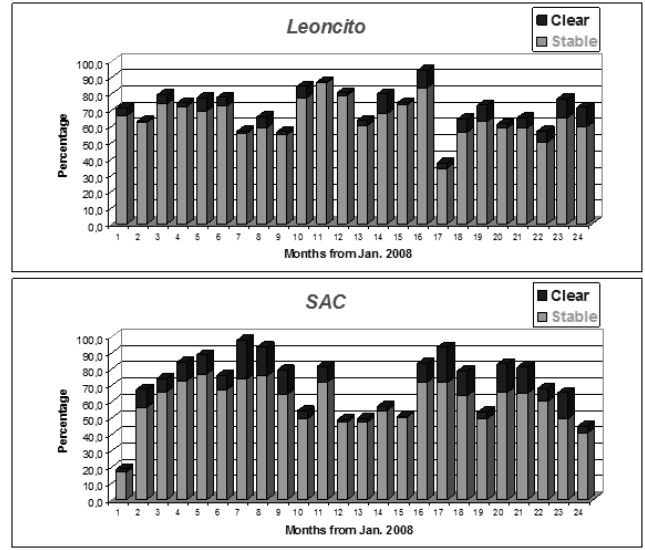}
  \caption{Distributions of stable and clear nights at Leoncito and SAC (2008-2009) obtained from {\bf{\it GOES 12}} satellite.}
             \label{stable1}
\end{figure}

\begin{figure}
  \centering
  \includegraphics[width=8.5cm]{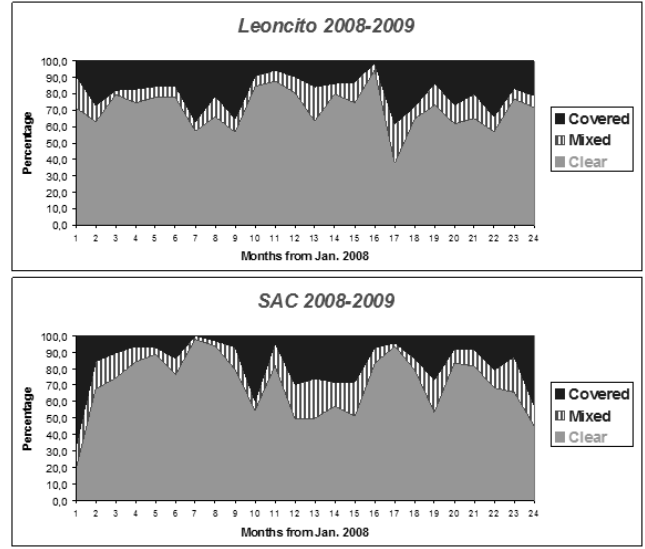}
  \caption{Distributions of clear, mixed and covered nights at Leoncito and SAC (2008-2009) obtained from {\bf{\it GOES 12}} satellite.}
             \label{clear1}
\end{figure}

In our analysis we found that, with regard to the cloud cover and atmospheric stability (Mean 2008-2009: 76.0\% Clear, 69.6\% Stable), the best site is SPM.
Then we have two sites to be considered on the same level: Iza$\tilde{n}$a and Leoncito (respectively, Mean 2008-2009: 68.6\% Clear, 62.6\% Stable. Mean 2008-2009: 70.6\% Clear, 64.9\% Stable).
El Leoncito has a uniform distribution of the observation time, less influenced by the seasons compared to the Iza$\tilde{n}$a. 
Iza$\tilde{n}$a, on the contrary has very stable summer months and lower quality winter months (especially in December and January).
We also noticed that the statistics of only two years may be influenced by a particular month: for example December 2009, SPM Clear 30.7\%; December 2009, Iza$\tilde{n}$a Clear 29.0\%; May 2009, Leoncito Clear 37.7\%; January 2008, SAC Clear 18.9\%.
Figures \ref{stable} and \ref{stable1} show the seasonal trends of the clear and stable nights for the sites in the northern hemisphere and southern hemisphere respectively.
In addition Figures \ref{clear} and \ref{clear1} show the seasonal trends of the clear, mixed and covered nights for the same sites.
The anomalous months and the characteristics described above in these figures are evident.
We have to make comments regarding the site of SAC (Mean 2008-2009: 70.0\% Clear, 59.7\% Stable).
SAC has generally a low cover, but it has bad months that are not always related to the seasons.
Furthermore, analysis of data showed strong fluctuations of the B3.
These fluctuations indicate the presence of large atmospheric airmass instabilities and may indicate 
strong winds as shown in Cavazzani et al. (\cite{cava10}).
In Sections \ref{swa} we have described an empirical procedure for the calculation of these winds (Mean 2008, strong wind $\Rightarrow 10m/s < v < 15m/s$, 12.9\%; extremely strong wind $\Rightarrow v > 15m/s$, 5.5\%. Mean 2009, strong wind , 10.8\%; extremely strong wind, 2.3\%).
This is consistent with the fact that SAC is the highest site (3600m).

\section{Conclusion}

Using {\bf{\it GOES 12}} satellite we have presented a homogeneous method in order to obtain the amount of available time fraction for four interesting sites for astronomical instrumentation. In order to have a complete analysis we have also collected meteorological characteristics from literature. In this analysis a wider spatial field is used in order to reduce the spatial noise: each value is the mean of $1^\circ\times1^\circ$ matrix.  The cloud coverage is obtained using the {\bf{\it GOES 12}} B4 band. Using the correlation of the three bands, B3, B4 and B6, we have computed an atmospheric correlation function as a further selection of the clear nights, and we have introduced the new concept of stable night. The years 2008 and 2009 are analysed. 

We found for the 2008 the following clear nights: 73.4\% Iza$\tilde{n}$a, 79.2\% at San Pedro Martir, 72.9\% Leoncito and 72.3 \% SAC. For the 2009 we found: 63.8\% Iza$\tilde{n}$a, 72.7\% at San Pedro Martir, 68.3\% Leoncito and 67.8 \% SAC. Instead, the number of stable nights for the 2008 is: 66.4\% Iza$\tilde{n}$a, 73.7\% at San Pedro Martir, 69.0\% Leoncito and 61.7 \% SAC. For the 2009 we found 58.8\% Iza$\tilde{n}$a, 65.4\% at San Pedro Martir, 60.8\% Leoncito and 57.7 \% SAC.

 \begin{table}
      \centering
      \begin{minipage}{200mm}
       \caption{Satellite available nights for the candidate sites. }
        \label{statN}
         \resizebox*{0.40\textwidth}{!} {
       \begin{tabular}{@{}lcccccccc@{}}
       \hline
                 &\multicolumn{4}{c}{Iza$\tilde{n}$a}&\multicolumn{4}{c}{SPM}  \\
                 &\multicolumn{2}{c}{Clear} & \multicolumn{2}{c}{Stable}&\multicolumn{2}{c}{Clear} & \multicolumn{2}{c}{Stable}\\
                 & 2008 & 2009 & 2008 & 2009 & 2008 & 2009 & 2008 & 2009 \\
                  \hline
   Mean           & 73.4  & 63.8 & 66.4 & 58.8  & 79.2 & 72.7 & 73.7 & 65.4\\
   \hline     
                 &\multicolumn{4}{c}{Leoncito}&\multicolumn{4}{c}{SAC}  \\
                 &\multicolumn{2}{c}{Clear} & \multicolumn{2}{c}{Stable}&\multicolumn{2}{c}{Clear} & \multicolumn{2}{c}{Stable}\\
                 & 2008 & 2009 & 2008 & 2009 & 2008 & 2009 & 2008 & 2009 \\
                 \hline
  Mean           &  72.9 & 68.3 & 69.0 & 60.8 & 72.3 & 67.7 & 61.7 & 57.7 \\ 
     \end{tabular}
           }
     \end{minipage}
     \end{table}

The mean of the two years gives a percentage of clear nights of 68.6\% Iza$\tilde{n}$a, 76.0\% at San Pedro Martir, 70.6\% Leoncito and 70.0 \% SAC, while the mean of the two years gives a percentage of stable nights of 62.6\% Iza$\tilde{n}$a, 69.6\% at San Pedro Martir, 64.9\% Leoncito and 59.7 \% SAC. \\
Iza$\tilde{n}$a shows a large variability of clear night in these two years, while the percentage of stable nights is about 6.0\% less that the clear nights.
SAC shows a large fluctuation in the considered years and the estimate of the percentage of clear nights does not take into account the wind speed (see Section \ref{swa}).

\subsection{ACKNOWLEDGMENTS}

This activity is supported by Strategic University of Padova funding by title "QUANTUM FUTURE".\\
Most of data of this paper are based on the CLASS (Comprehensive Large Array-data Stewardship System).\\ CLASS is an electronic library of NOAA environmental data.\\ This Web-site provides capabilities for finding and obtaining those data, particularly NOAA's Geostationary Operational Environmental Satellite data.

\label{lastpage}


\begin{thebibliography}{}

\bibitem[\protect\citeauthoryear{2007}{}]{giova07}
Acker, J. G. and G. Leptoukh, 2007, Eos Trans. AGU, 88, p.14.

\bibitem[\protect\citeauthoryear{2011}{}]{cava11}
Cavazzani, S., Ortolani, S., Zitelli, V., 2011, MNRAS, 419, 3080

\bibitem[\protect\citeauthoryear{2010}{}]{cava10}
Cavazzani, S., Ortolani, S., Zitelli, V., Maruccia,Y. 2010, MNRAS, 411, 1271

\bibitem[\protect\citeauthoryear{2011}{}]{dee}
Dee et al., 2011, Q. J. R. Meteorol Soc., 137, p.553 

\bibitem[\protect\citeauthoryear{2010}{}]{dv10}
Della Valle,A., Maruccia,Y., Ortolani, S., and Zitelli,V., 2010, MNRAS, 401,1904 

\bibitem[\protect\citeauthoryear{2002}{}]{era02}
Erasmus,D., and Sarazin, M., 2002, Vernin J. et al. Eds., Asp. Conf. Series 266, p.310.

\bibitem[\protect\citeauthoryear{2001}{}]{era2001}
Erasmus,D.,\& Maartens D., 2001, Final Report to ESO 2001(58311/ODG/99/8362/GWI/LET; garching ESO)

\bibitem[\protect\citeauthoryear{2006}{}]{era06}
Erasmus,D., van Rooyen,R., 2006, in Stepp, L.M. Ed., Proc. of SPIE Vol. 6267, Ground-based and Airborne Telescopes.


\bibitem[\protect\citeauthoryear{2005}{}]{graham}
Graham, E., Sarazin, M., Beniston, M., Collet, C., Hayoz, M., Neun, M. and P. Casals, 2005, Meteorological Applications, 12, pp 77-81.

\bibitem[\protect\citeauthoryear{2008}{}]{graham1}
Graham,E., Sarazin,M., Kurlandczyk,H., Neun,M., Mätzler,C., ,2008, Proceedings of the SPIE, 7012 (69), 10pp.

\bibitem[\protect\citeauthoryear{2006}{}]{lombardi06}
Lombardi,G.,Zitelli,V., Ortolani,S., Pedani,M J., 2006, PASP, 118,1198 

\bibitem[\protect\citeauthoryear{1985}{}]{murdin85}
Murdin, P., 1985, Vistas in Astronomy, 28, 449

\bibitem[\protect\citeauthoryear{1984}{}]{osmer84}
Osmer, P.S., Woos,H.J., 1984, ESOC. 18, 950

\bibitem[\protect\citeauthoryear{2008}{}]{rovero}
Rovero A.C., Romero G.E., Allekotte I., Beton X., Colombo 
E., Etchegoyen A., Garciaa B., Garcia-Lambas D., Levato H., Medina M.C., 
Muriel H. and Recabarren P., 2008, arXiv 0810.0628v1.


\bibitem[\protect\citeauthoryear{2010}{}]{san10}
Sanroma, E., Palle, E., Sanchez-Lorenzo, A., 2010, Environ. Res. Lett. 5 1-6

\bibitem[\protect\citeauthoryear{2006}{}]{sara}
Sarazin, M., Graham, E. and Kurlandczyk, H., 2006, The Messenger, 125, pp 44-47.

\bibitem[\protect\citeauthoryear{2007}{}]{tapia07}
Tapia, M., Hiriart, D., Richer, M., and Cruz-Gonzalez, I., 2007, RevMexAA (Series Conf.), 31, 47.

\bibitem[\protect\citeauthoryear{1992}{}]{tapia92}
Tapia, M., 1992, RevMexAA, 24, 179-186.


\bibitem[\protect\citeauthoryear{2008}{}]{varela}
Varela, A.M., Bertolin, C., A., Mu$\tilde{n}$oz-Tu$\tilde{n}$on, C.,Ortolani, S., and Fuensalida, J.,J., 2008, MNRAS 391, 507-520.





\end{thebibliography}
\end{document}